\documentclass[11pt]{article}

\usepackage[margin=1in]{geometry}
\usepackage{tipa} 
\usepackage{indentfirst}
\usepackage{xpatch}
\usepackage[font=small, labelfont=bf]{caption}
\usepackage[scaled]{helvet}
\usepackage{courier}

\normalfont
\usepackage[T1]{fontenc}
\usepackage{multirow}
\usepackage{CJKutf8}

\long\def\symbolfootnote[#1]#2{\begingroup%
\def\thefootnote{\fnsymbol{footnote}}\footnote[#1]{#2}\endgroup}
\def\blfootnote{\xdef\@thefnmark{}\@footnotetext}

\usepackage{graphicx}	
\usepackage{color}
\usepackage{subfigure}
\usepackage{booktabs}
\usepackage[flushleft]{threeparttable}
\usepackage{float}
\usepackage{sidecap}
\usepackage{amssymb, bm, mathrsfs}
\usepackage{amsfonts, amsmath}
\usepackage{mathpazo}
\normalfont

\usepackage{paralist}
\usepackage{url}
\usepackage[hyperfootnotes=false]{hyperref}
\usepackage[semicolon, sort&compress]{natbib}
\usepackage{seqsplit}

\usepackage{algorithm}
\usepackage{algpseudocode}
\usepackage{authblk}

\renewenvironment{thebibliography}[1]{%
  \begin{oldthebibliography}{#1}%
  \small}%
  {\end{oldthebibliography}}

\title{\bf{DrugLLM: Open Large Language Model for Few-shot Molecule Generation}}

\author[1,4]{Xianggen Liu}
\author[1,4]{Yan Guo}
\author[1]{Haoran Li}
\author[2]{Jin Liu}
\author[1,3]{Shudong Huang}
\author[2]{Bowen Ke}
\author[1]{Jiancheng Lv}

\affil[1]{College of Computer Science, Sichuan University, Chengdu 610065, China}
\affil[2]{Department of Anesthesiology, Laboratory of Anesthesia and Critical Care Medicine, National-Local Joint Engineering Research Centre of Translational Medicine of Anesthesiology, Frontiers Science Center for Disease-related Molecular Network, West China Hospital, Sichuan University, Chengdu 610041, China}
\affil[3]{Information Technology Research Center, Beijing Academy of Agriculture and Forestry Sciences, Beijing 100097, China}
\affil[4]{These authors contributed equally}

\date{}		

\begin{document}
\maketitle

\begin{abstract}
Large Language Models (LLMs) have made great strides in areas such as language processing and computer vision. Despite the emergence of diverse techniques to improve few-shot learning capacity, current LLMs fall short in handling the languages in biology and chemistry. For example, they are struggling to capture the relationship between molecule structure and pharmacochemical properties. Consequently, the few-shot learning capacity of small-molecule drug modification remains impeded. In this work, we introduced DrugLLM, a LLM tailored for drug design. During the training process, we employed Group-based Molecular Representation (GMR) to represent molecules, arranging them in sequences that reflect modifications aimed at enhancing specific molecular properties. DrugLLM learns how to modify molecules in drug discovery by predicting the next molecule based on past modifications. Extensive computational experiments demonstrate that DrugLLM can generate new molecules with expected properties based on limited examples, presenting a powerful few-shot molecule generation capacity.

\end{abstract}


\section{Introduction}

Small molecules play a critical role in drug discovery due to their ability to bind to specific biological targets and modulate their functions\citep{scott2016small, offensperger2024large, kovachka2024small}. In fact, over the past decade, according to the approval records of the U.S. Food and Drug Administration (FDA), small molecule drugs have accounted for 76\% of the total number of drugs approved for the market \citep{brown2021decade, norsworthy2024fda}. Small molecules are advantageous in drug discovery because they can be synthesized relatively easily and have good bioavailability, making them more likely to reach their intended targets in vivo (especially when passing through cell membranes) \citep{{vargason2021evolution}}. However, based on the current research technologies, it is very challenging to design a molecule with ideal properties, and it will consume a lot of resources and time. For example, in the drug development process, it takes 9 to 12 years and billions of dollars to find an effective drug \citep{adams2006estimating, dickson2009cost}. 

The vastness of the search space for new molecules, with up to $10^{60}$ synthetically creatable drug-like molecules, presents a significant challenge in drug design as chemists must navigate this immense space to identify molecules that interact with a biological target, and while modern techniques allow for the testing of over $10^6$ molecules in a laboratory setting, larger experiments become prohibitively expensive \citep{segler2018generating}. As such, computational tools are necessary to help narrow down the search space. Virtual screening is one such strategy used to identify promising molecules from millions of existing or billions of virtual molecules \citep{lyu2023modeling}. But high-throughput screening and virtual screening only consider known molecules that are synthetically accessible and fall short of producing novel molecules \citep{crunkhorn2022screening, sadybekov2022synthon}.

As an alternative to exploring the huge molecule space, de novo drug design exhibits the remarkable ability to generate entirely novel and distinctive molecules \citep{ren2024small, tropsha2024integrating}. Traditional de novo drug design methods use molecular construction rules to generate new molecules based on the receptor structure~\citep{waszkowycz1994pro_ligand, gillet1995sprout} or the ligand structure~\citep{afantitis2011ligand}. Recently, deep learning and reinforcement learning techniques offer promising potential in de novo drug design due to their powerfutl approximation capacity \citep{li2018, blaschke2018, putin2018, liu2020chance}. In particular, \citet{popova2018deep} integrates both generative and predictive neural networks to generate novel targeted chemical libraries in a trial-and-error manner. \citet{jin2018junction} propose a junction tree-based variational autoencoder to learn a continuous molecule space and generate new molecules by sampling from it. 

Despite the development of various deep learning (DL)-based methods for de novo drug design, the field of few-shot molecule generation remains notably underexplored. Few-shot molecule generation aims to generate new molecules with expected properties given limited molecule examples. Most of the current de novo drug design approaches require thousands of data for learning \citep{korshunova2022generative}. However, data scarcity is a prevalent issue in drug discovery due to the high costs associated with biological experimentation \citep{wang2023multitask}. Consequently, the ability to perform few-shot generation is of paramount importance for the advancement of de novo drug design techniques.

The large language models (LLMs) have achieved significant progress in natural language processing, especially in the few-shot learning problem \citep{ahmed2022few}. Despite the emergence of diverse LLMs \citep{chang2024survey}, they fall short in handling the languages in biology and chemistry \citep{ross2022large}. For example, they are still struggling to capture the relationship between molecule structure and the corresponding properties. Therefore, how do we build an LLM that can accurately characterize the "structure-effect-metabolism-toxicity" relationships in molecules?

In this work, we presented DrugLLM, a large language model for drug design. In DrugLLM, molecules are represented using Group-based Molecular Representation (GMR), which is a novel type of molecular representation to address token abundance, cyclic complexity, and structural sensitivity inherent in SMILES\citep{o2012towards}. GMR makes structural groups as units to build the topological structure of molecules and transform them into linear sequences. 

Furthermore, we will provide an in-depth exposition of the training methodology employed by DrugLLM. The methodology organizes the modification sequences in accordance with specific molecular properties. Drawing an analogy, a case of molecule modification (a pair of molecules with similar structures) toward a certain property serves as a ''sentence''. Multiple cases of modification toward the identical property constitute a paragraph. By continuously predicting the next molecule based on the modification history, DrugLLM learns the intrinsic relationship between the molecular structure and the corresponding property. To the best of our knowledge, DrugLLM is the first large language model for few-shot molecule generation.

\section{Results}

\subsection{The DrugLLM Framework}
The focus of this work is to train a large language model that can capture the relationship between the molecule structure and the corresponding chemical and biological activities. Unlike ChatGPT \citep{brown2020language} and LLaMA \citep{touvron2023llama}, which are trained on massive text data from the internet, and DrugGPT \citep{li2023druggpt}that uses SMILES as its representation, DrugLLM employs Group-based Molecular RepresSMILESentation (GMR) strings as its primary language representation.

GMR leverages structural groups to depict molecular architectures, thereby effectively overcoming three principal challenges inherent in the application of SMILES notation within model processing contexts: (1) \textbf{Token Abundance}: In the SMILES format, each character is considered a separate token, which can result in an unwieldy number of tokens and subsequently consume considerable computational resources during training. (2) \textbf{Cyclic Complexity}: The representation of cyclic structures within molecules is particularly intricate in SMILES, which increases the difficulty of model training. (3) \textbf{Structural Sensitivity}: Even minor alterations in a molecule's structure can produce significant discrepancies in the corresponding SMILES representation.

\begin{figure}
\centering
\includegraphics[width=0.94\linewidth]{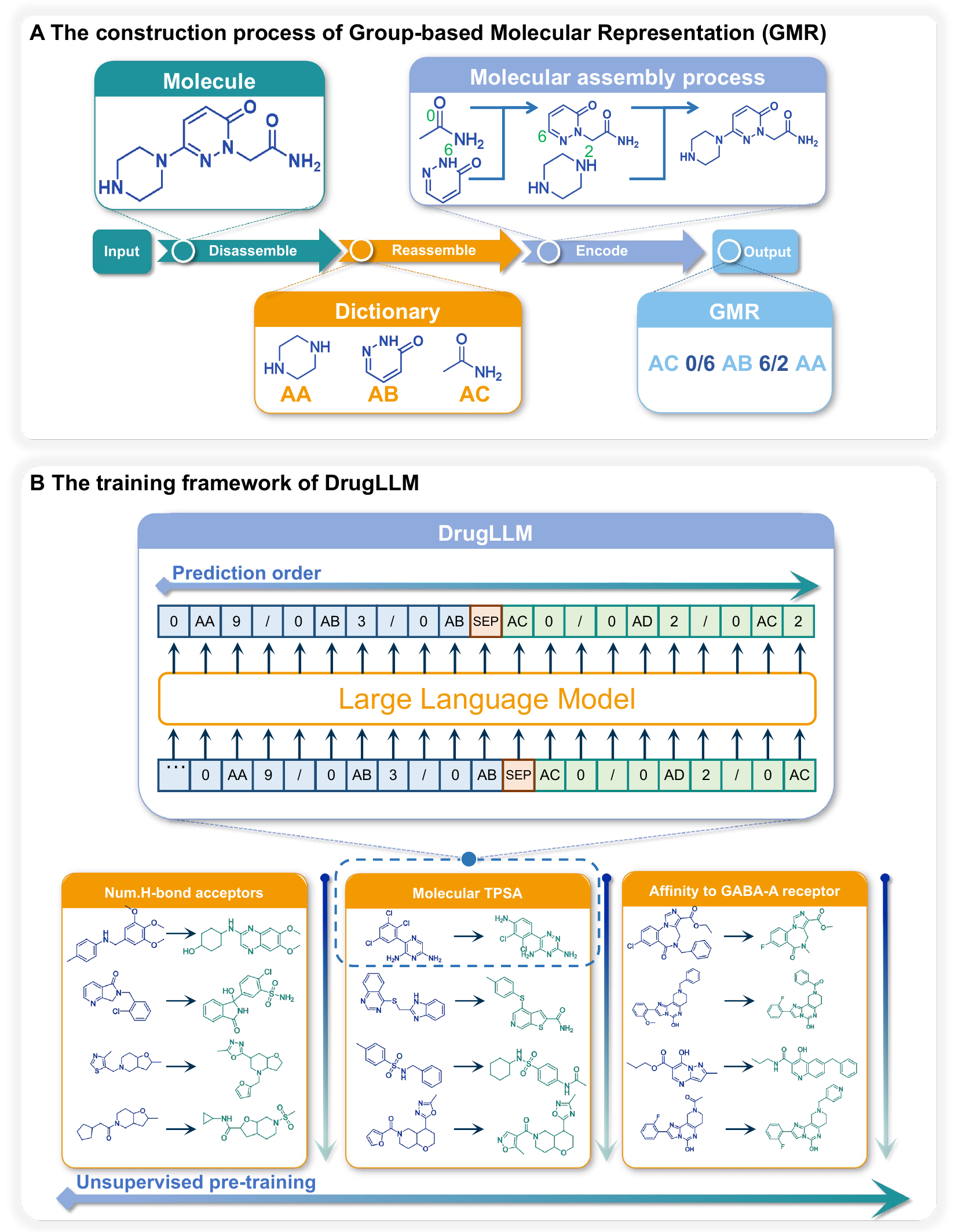}
\caption{Schematic overview of the DrugLLM framework. \textbf{A}, Construction process. Group-based Molecular Representation (GMR) is constructed from molecular structure units. \textbf{B}, Training framework. DrugLLM is trained on molecular modifications, with each paragraph representing a unique attribute. Each paragraph is self-contained and represents multiple characteristics, with different paragraphs corresponding to different attributes.}
\label{fig:model}
\end{figure}

As shown in Figure~\ref{fig:model}A, the GMR framework employs unique string identifiers to represent distinct structural groups. These identifiers are linked by numerical position data flanked by a slash. By employing GMR, the model can recognize molecular strings by treating structural groups as units, thereby reducing the number of input and output tokens. Additionally, GMR removes cyclic structures by merging them, simplifying the logic of molecular assembly and lowering the difficulty of model recognition. It also minimizes the discrepancies in SMILES strings that result from even small structural changes.

To train DrugLLM, we construct sentences and paragraphs composed of molecule modifications as training data (Figure~\ref{fig:model}B). Specifically, DrugLLM regards the modification between two molecules with similar structures as a sentence. A series of such modifications are viewed as a sequence of sentences that form a paragraph. Note that we impose the constraint that the molecular modifications in a paragraph should characterize the identical property. For instance, if the first three cases of molecule modifications describe the increase in the number of hydrogen bond acceptors, we expect all subsequent sentences in that paragraph to also discuss the increase in acceptor numbers. In this way, the contents of a paragraph are concentrated, making DrugLLM able to auto-regressively predict the next token based on the previous contexts. Moreover, each paragraph exhibits autonomy, encompassing a diverse array of molecular characteristics. The distinct paragraphs engage with unique molecular properties, necessitating that DrugLLM be endowed with the capability for in-context learning (a form of few-shot learning).

However, there are few related datasets available. In this work, we collect the tabular form of the molecule datasets from the ZINC database \citep{wu2018moleculenet} and the ChEMBL platform \citep{mendez2019chembl, davies2015chembl}, and convert them into the corresponding sentences and paragraphs. In total, we collected over 25, 000, 000 modification paragraphs and 200, 000, 000 molecules to build the training dataset (Table \ref{tab:dataset}). The dataset involves over 10,000 different molecular properties or activities, such as the count of hydrogen bond acceptors and the topological polar surface area (TPSA). Considering that the few-shot learning capability of machine learning models arises from their exposure to a sufficient variety of training tasks, a large scale of different paragraphs could enforce DrugLLM captures the intrinsic nature of molecule design in a few-shot fashion.

\begin{table}
	\caption{Pre-training data of DrugLLM. The training data include ZINC and ChEMBL databases. Similar molecules are collected together to build the modification paragraphs. A modification paragraph contains multiple molecule modifications that aim to improve or decrease the same molecular properties.} 
	
	\begin{center}
		\begin{tabular}{lcccccccccc}
			\hline\noalign{\smallskip}
			Dataset & \# of molecules & \# of paragraphs & \# of tasks & Disk size  \\
			\hline\noalign{\smallskip}
			ZINC & 4.5M & 0.6M & 770 & 780M \\
			ChEMBL & 180.2M & 24M & 10100 & 30G \\
			\hline
		\end{tabular}
	\end{center} 
	\label{tab:dataset}
\end{table}

Following recent work on the pre-training large language models, DrugLLM is based on the Transformer architecture \citep{vaswani2017attention}. We adopt the parameters of LLama 7B \citep{touvron2023llama} and expand the vocabulary by introducing the frequently-used SMILES tokens. These tokens are divided by the byte pair encoding (BPE) algorithm \citep{sennrich-etal-2016-neural}. We train DrugLLM using the AdamW optimizer for six weeks on eight NVIDIA RTX 3090 GPUs, where it learns to generate the paragraphs from scratch. In the view of machine learning, the paragraph plays as a process of few-shot molecule generation. Therefore, the trained DrugLLM is able to directly perform few-shot molecule generation without further fine-tuning. 

\subsection{DrugLLM is a few-shot learner in molecule optimization toward physicochemical properties}

\begin{figure}
\centering
\includegraphics[width=0.94\linewidth]{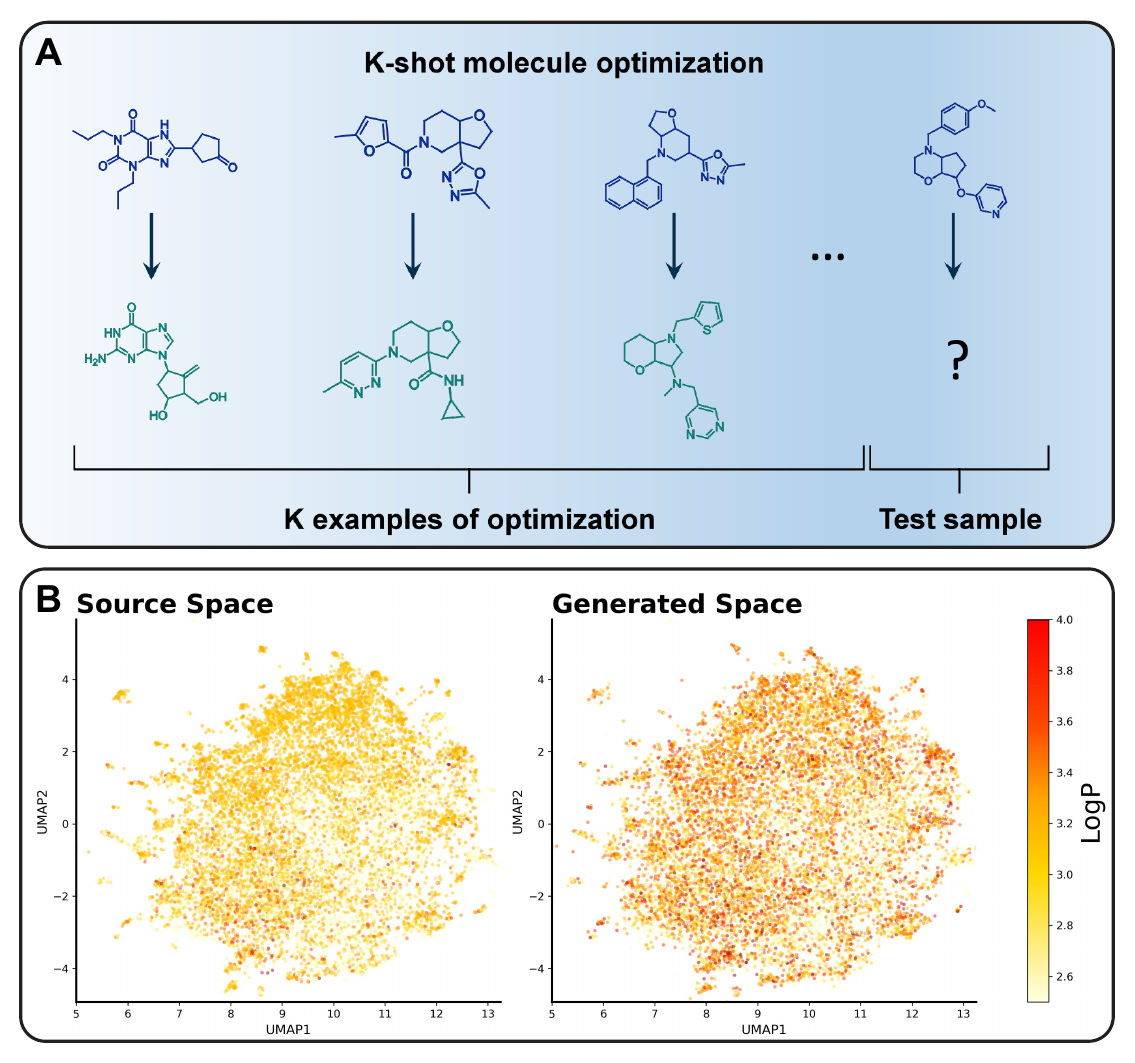}
	\caption{Visualization of few-shot molecule optimization. \textbf{A}, The training and testing examples of few-shot molecule optimization. \textbf{B}, Chemical space navigation by transfer learning. The UMAP plot shows the distribution of 15000 molecules selected from the source space and their corresponding 15000 molecules in the generated space after LogP property optimization. Different values of LogP are represented by different colors.}
	\label{fig:Umap}
\end{figure}

Figure~\ref{fig:Umap}A illustrates our approach under the $K$-shot learning framework, where we provide the model with $K$ pairs of example modifications and a benchmark molecule. The objective of the model is to generate a new molecule that not only maintains structural similarity to the benchmark molecule but also exhibits superior properties (either increased or decreased, as guided by the examples). Due to input token limitations, we restrict the number of molecular optimization examples to a maximum of nine pairs. To visually represent the structural similarity between the benchmark and generated molecules in the chemical space, we employ the Uniform Manifold Approximation and Projection (UMAP) method to create a chart (Figure~\ref{fig:Umap}B). There is a high degree of consistency between the distribution of the generated molecules (on the left) and the source molecules (on the right). This distributional similarity, coupled with the notable improvement in the LogP properties of the generated molecules, underscores the robust capability of the model to optimize the properties of the benchmark molecule.

To evaluate the capacity of DrugLLM in terms of few-shot molecule generation, we select four physicochemical properties that are not seen by DrugLLM as the test tasks, including the water-octanol partition coefficient (LogP), solubility, synthetic accessibility, and topological polar surface area (TPSA). As these four molecular properties can be accurately estimated by machine learning-based scripts, they are widely used in the assessment of molecule generation models. For comparison, we take the junction tree-based variational auto-encoder (JTVAE) \citep{jin2018junction}, the variational junction tree neural network (VJTNN) \citep{jin2020hierarchical}, and the scaffold-based molecule generator (MoLeR) \citep{maziarz2021learning}. We also include a random generation control implemented by random sampling based on the latent space of JTVAE. The quality of the generated molecules was assessed based on their success rate and molecular similarity. The success rate represents the proportion of generated molecules that adhere to the rule of examples of modifications. To avoid the generation bias of the generators, the input contexts (i.e., the prompts of the language models) describe the increment or decrement of the property with a balanced proportion. Although these models are not initially designed for few-shot learning, they are state-of-the-art molecule generators in literature.

\begin{figure}
\centering
\includegraphics[width=0.94\linewidth]{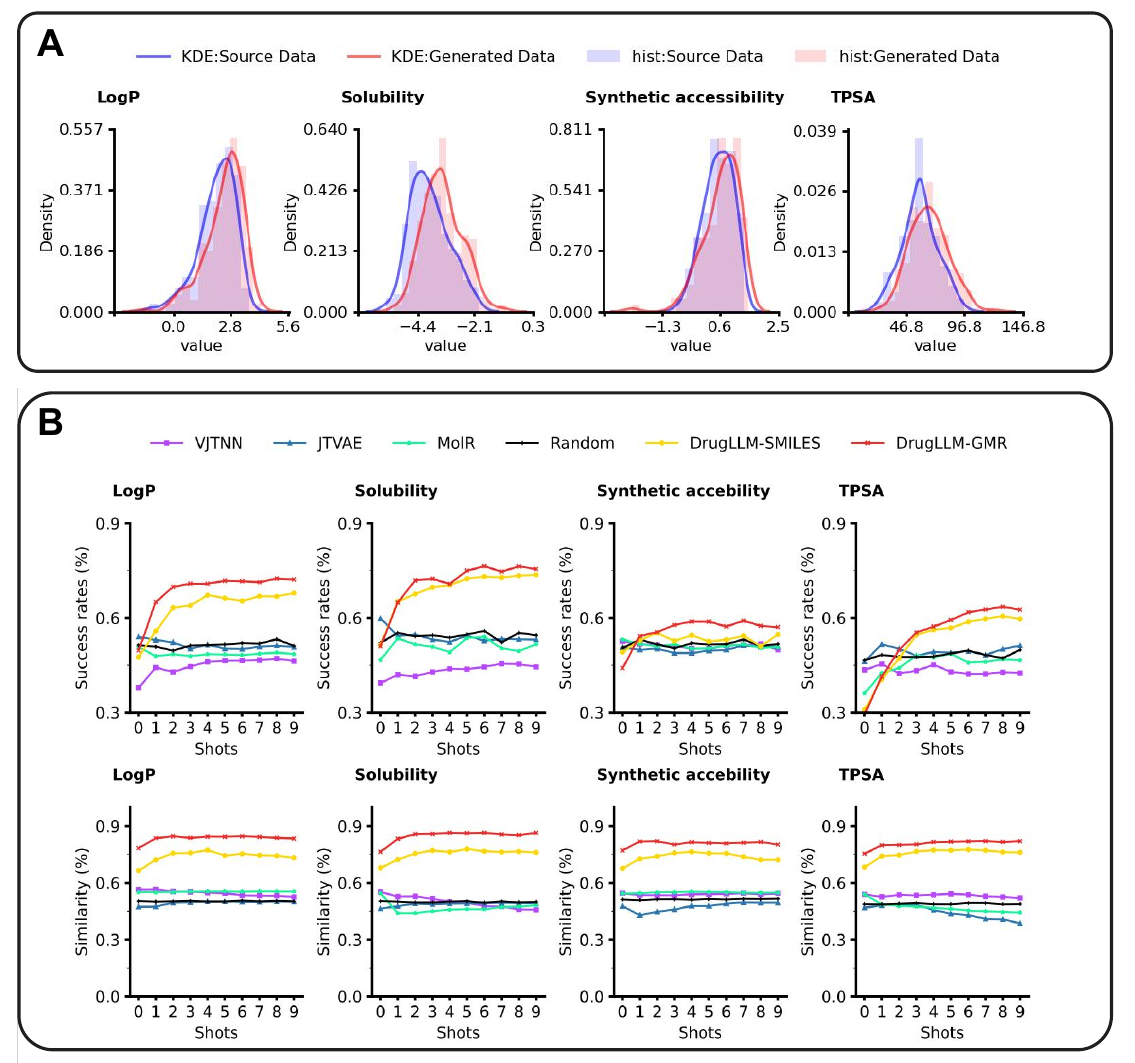}
	\caption{The performance of few-shot molecule optimization in physiochemical properties. \textbf{A}, The distribution of the water-octanol partition coefficient (LogP), Solubility, Synthetic Accessibility, and Topological Polar Surface Area (TPSA) for the source data and the generated data, represented by Kernel Density Estimation (KDE). The KDE demonstrations of the source data and the generated data are displayed by blue solid lines and red solid lines, respectively. Histograms (hist) for the source data and the generated data are represented by blue bars and red bars, respectively. \textbf{B}, The performance of the generation methods in terms of the success rates and the generation similarities.}
	\label{fig:prep}
\end{figure}

We first present the distributions of several key properties - LogP, solubility, synthetic accessibility, and TPSA - for both the source and generated data in Figure~\ref{fig:prep}A. These distributions are visualized using Kernel Density Estimation (KDE). The significant numerical improvements of the model in optimizing these key properties are clearly demonstrated, further attesting to the effectiveness of the model in optimizing molecular properties.
Next, as shown in Figure~\ref{fig:prep}B, we report the performance of few-shot generation with respect to the LogP value. We note that the three baseline molecule generators, namely JTVAE, VJTNN, and MoLeR, just obtained a success rate of about 50\%, which is similar to a random generation. In contrast, DrugLLM exhibits a progressive improvement in few-shot molecule generation, with the accuracy of the generated molecules increasing incrementally to 75\% as the number of shots increases. Performance comparisons on molecular solubility, synthetic accessibility, and TPSA are similar and consistent. When it comes to similarity, it is typically more challenging to optimize a molecule with fewer modifications (i.e., higher similarity). Despite this, DrugLLM maintains a higher success rate even with increased generation similarity, underscoring its superior performance in the few-shot generation. Furthermore, we note that DrugLLM-GMR slightly outperforms DrugLLM-SMILES, highlighting the benefits of GMR in large model training.

\subsection{DrugLLM is a few-shot learner in molecule optimization towards biological activities}
Since DrugLLM shows impressive few-shot generation capacity in physicochemical properties, we next validate the effectiveness of DrugLLM in the biological activities of molecules, which is more challenging. The molecules produced by DrugLLM are usually novel and are not recorded in the ChEMBL database. Unlike the physicochemical properties mentioned above, the biological activities of molecules (e.g. the Ki value on streptokinase A) are difficult to estimate by chemical or physical rules. In addition, the lengthy time and high costs associated with wet-lab experiments hinder the large-scale evaluation of molecules. Instead, we leverage a message-passing neural network to predict biological activities. Specifically, before building the DrugLLM dataset through the ChEMBL database, we scan all biological activities and select the one that has relatively adequate samples ($N\ge800$) and could induce an accurate property predictor (Pearson $r\ge0.75$). Finally, 10 activities are obtained and excluded from the training data. These activities are used to test the optimization of the few-shot DrugLLM. As the Pearson correlation of the prediction of the predictor is greater than 0.75, it could correlate well with the real evaluation in statistics.

\begin{table}
\caption{The performance of few-shot molecule optimization toward biological activities} 
\begin{center}
    \resizebox{0.97\linewidth}{!}{
		\begin{tabular}{lcccccccccc}
			\hline\noalign{\smallskip}
			\multirow{1}{*}{Target} &
						\multirow{1}{*}{Property} &
			\multirow{1}{*}{Random} & \multicolumn{1}{c}{JTVAE} &  \multicolumn{1}{c}{VJTNN} &  \multicolumn{1}{c}{MoLeR} &   \multicolumn{1}{c}{DrugLLM}  \\
			\hline
			\noalign{\smallskip}
   		cAMP-dependent protein kinase alpha-catalytic subunit   & Ki & 0.58 &0.55 &0.42 &0.51 &\textbf{0.69} \\ 
      	Cyclin-dependent kinase 2   & Ki & 0.51 &0.50 &0.53 &0.50 &\textbf{0.63} \\ 
        	Cyclin-dependent kinase 5   & Ki & 0.49 &0.54 &0.57 &0.57 &\textbf{0.65} \\ 
			Dual-specificity tyrosine-phosphorylation regulated kinase 1A   & Ki &0.53 &0.54 &0.38 &0.52 &\textbf{0.66} \\ 
			Rho-associated protein kinase 1   & Ki &0.54 &0.59 &0.48 &0.52 &\textbf{0.76} \\ 
			Rho-associated protein kinase 2   & Ki & 0.55 &0.57 &0.41 &0.54 &\textbf{0.72} \\ 
			Serine/threonine-protein kinase Aurora-A   & Ki & 0.48 &0.53 &0.44 &0.54 &\textbf{0.60} \\ 
            Serine/threonine-protein kinase Aurora-B   & Ki & 0.53 &0.60 &0.55 &0.59 &\textbf{0.74} \\ 
			Serine/threonine-protein kinase MST2   & Ki & 0.47 &0.58 &0.39 &0.55 &\textbf{0.76} \\ 
			Vascular endothelial growth factor receptor 2   & Ki & 0.43 &0.44 &0.42 &0.48 &\textbf{0.59} \\ 
			Average   & -  & 0.51 &0.54 &0.46 &0.53 &\textbf{0.68} \\ 
			\hline
		\end{tabular}
  }
	\end{center} 
	\label{tab:bio}
\end{table}

We observe that the three generator baselines fail to obtain meaningful improvement compared with the random generation (Table \ref{tab:bio}), indicating that these molecule generators still struggle to capture the modification rules underlying the limited examples. As for DrugLLM, it significantly outperforms the other baselines by a large margin in most of the test properties. In particular, DrugLLM can generate appropriate molecules that bind to Rho-associated protein kinase 1, with a success rate of 76\%. Note that these test properties are not observable in the training of DrugLLM. These results demonstrate that DrugLLM is able to figure out the intrinsic rules of the molecule modifications given a limited number of examples for an unknown molecular property.

\subsection{DrugLLM support instruction guided molecule optimization in a zero-shot manner}
Previous experiments demonstrated that DrugLLM can accept multiple pairs (i.e., shots) of molecules as references to learn the modification rules and generate new molecules with the desired properties. In this section, we explore zero-shot molecule optimization, which involves generating modified molecules with better properties of interest according to natural language instruction without any specific training instances. In this experimental setting, we assume that when DrugLLM is trained on a large scale of properties and their compositions, it is capable of generalizing to optimize the molecules toward unseen compositions of properties. Therefore, we move six optimization tasks out of the DrugLLM training set in advance and leave them as test tasks. For example, all samples related to the optimization of quantitative estimation of drug similarity (QED) and topological polar surface area (TPSA) are in the test set, but the samples related to the optimization of each single property are in the training set. Based on this setting, we build the test set that contains 6, 000 instructions, each optimization task for 1000. The generated molecules are evaluated by the Python scripts via the RDKit library.

It is noteworthy that DrugLLM is one of the very few approaches that support zero-shot molecule optimization. Apart from ChatGPT\citep{brown2020language}, GPT-4\citep{achiam2023gpt}, and ChatGLM\citep{zeng2022glm}, the other large language models, e.g., LLaMA\citep{touvron2023llama}, Alpaca\citep{maeng2017alpaca}, and Vicuna\citep{chiang2023vicuna}, were unable to generate valid SMILES strings of molecules. Thus, we only compared the zero-shot molecule optimization capacity between DrugLLM and ChatGPT, GPT-4, and ChatGLM, all of which are trained on thousands of billion tokens. The challenge of zero-shot molecule optimization lies in two folds. On the one hand, the mapping between semantics and molecular property is hard to learn from the general corpus. On the other hand, the biological data related to the structure-property relationship is not sufficient due to the lengthy time and high costs associated with wet-lab experiments. As a result, we observed that ChatGLM struggled to provide appropriate molecules on all the zero-shot molecule optimization tasks (Table~\ref{tab:zero}), with most generations outputting the duplicated molecules with the input ones. In addition, ChatGPT and GPT-4 were able to understand the instructions and optimize some of the given molecules. However, the success rate is relatively low. In terms of DrugLLM, it improves the optimization success rates by significant margins compared with the other LLM, indicating a better capacity in instruction understanding and molecule optimization.

\begin{table}
	\caption{The success rates (\%) of individual methods in zero-shot molecule optimization.} 
	
	\begin{center}
		\begin{tabular}{lcccccccccc}
			\hline\noalign{\smallskip}
			\multirow{1}{*}{Method} & ChatGLM &
			\multirow{1}{*}{ChatGPT} & \multicolumn{1}{c}{GPT-4} & \multicolumn{1}{c}{DrugLLM}  \\
			\hline
			\noalign{\smallskip}
			QED \& FractionCSP3     & 0.02  & 0.11 & 0.20 & 0.40 \\ 
			QED \& TPSA & 0.03 & 0.15 & 0.20 & 0.47  \\ 
			QED \& \# Rotatable bonds & 0.06 & 0.15 &0.10 &0.59\\
			LogP \& FractionCSP3 & 0.03 & 0.30 & 0.40 &0.60\\
			LogP \& TPSA & 0.04 & 0.19 & 0.43 & 0.55  \\ 
			LogP \& \# Rotatable bonds & 0.04 & 0.19 &0.05 &0.61\\
			\hline
		\end{tabular}
	\end{center} 
	\label{tab:zero}
\end{table}

\section{Discussion}
In this study, we introduce a computational task named few-shot molecule optimization, which is one of the few-shot generation problems. Given a molecule of interest, the task involves generating new molecules that adhere to the rules underlying the few modification examples. Although various few-shot learning tasks have been proposed and investigated \citep{stanley2021fs,ma2021few,zhang2024human}, there are few studies that consider few-shot molecule generation. Few-shot molecule optimization requires the model to capture the abstract rules in a small number of examples and apply the rules to new molecules, requiring a comprehensive understanding of ''structure-effect-metabolism-toxicity'' relationships. As expected, the current methods struggle to accomplish few-shot molecule optimization, including ChatGPT and the other competitive molecule generators. In comparison, DrugLLM exhibits impressive performance, taking a solidified step toward general artificial intelligence in drug design.

DrugLLM is a large language model (LLM), built on a large number of textual data that span a wide variety of small molecules and biological domains. Recently, LLMs, such as ChatGPT\citep{brown2020language}, Alpaca \citep{maeng2017alpaca}, and ChatGLM \citep{zeng2022glm}, have amazing capabilities for general-purpose natural language generation. However, they are designed for general use and lack profound biological and pharmaceutical knowledge. We notice that there are also several LLMs for biological and medical fields, such as BioGPT \citep{luo2022biogpt} and DrugGPT \citep{li2023druggpt}. However, these LLMs still follow traditional training strategies that learn to generate natural language as in the biomedical article. This raises an open question that how LLM understands the language of biology and chemistry and how to perform few-shot learning in this field. In this work, we propose a novel solution in which DrugLLM iteratively performs similar molecule modifications according to the context using GMR. Experiments demonstrated its exceptional effectiveness in a few-shot molecule optimization.

Despite the advantages of our method, this study has several limitations. Firstly, DrugLLM merely supports up to 9 shots of molecule modifications due to the limitation of the hardware. In such an input length, we have validated the few-shot learning capacity of DrugLLM. In the future, we will increase the input length (i.e., the number of shots) to achieve more impressive performance in drug design. Secondly, the zero-shot molecule optimization of DrugLLM is relatively elementary. The current DrugLLM is only capable of optimizing the molecules toward the composition of two known molecular properties. The zero-shot learning ability for arbitrary instructions is still lagging behind. Thirdly, the GMR currently in use has difficulty representing a small number of complex molecules under certain circumstances and lacks certain standardization measures. In the future, we will optimize for special cases of GMR representation and add standardization methods to reduce the small number of encoding errors that the model may generate.

In conclusion, this study presents the first attempt to build a large language model for few-shot molecule generation and optimization. Based on the tabular data related to molecular properties and biological activities, we build a large-scale textual corpus in the format of the sequences of molecule modifications. DrugLLM is trained to predict the next molecules based on historical modifications in an autoregressive manner. In extensive computational experiments, we observed that DrugLLM surpassed all the competitive methods (including GPT4) in optimizing new molecules in the few-shot learning setting on over 20 properties or biological activities. These results establish the substantial enhancement of efficacy facilitated by our proposed methodology in molecule generation and optimization, highlighting the potential of DrugLLM as a powerful computational tool in drug discovery.

\section{METHOD DETAILS}
\subsection{Data collection and preparation}
To train and analyze the DrugLLM model, we construct a large-scale dataset from the ZINC \citep{wu2018moleculenet} and ChEMBL \citep{mendez2019chembl, davies2015chembl} datasets. ZINC is a free database that contains more than 230 million purchasable compounds in ready-to-dock, 3D formats. We filter the druglike molecules from ZINC and obtain 4.5 million molecules. ChEMBL is a comprehensive repository for bioactive compounds with their properties. We gather bioactivity data from the ChEMBL database with the corresponding Web resource client. Following the preprocessing pipeline in \citep{stanley2021fs}, we excluded all compounds that are not drug-like molecules. A standard cleaning and canonicalization procedure was applied to filtered compounds. All of the molecules were represented by SMILES strings and labeled with certain properties. To facilitate property comparison between two molecules, we only consider property categories with real numbers. Therefore, we obtained a large-scale dataset that comprised thousands of tabular data, each table corresponding to hundreds of molecules measured by an identical property.

Based on the collected tabular data, we then transform them into meaningful textual sentences and paragraphs. In particular, we regard the modification between two molecules with similar structures as a sentence and multiple cases of molecular modifications as a paragraph. In the meantime, we stipulate that the molecular modifications in a paragraph should describe the same property changes. In other words, if the first two cases of molecule modifications indicate the increase of solubility, we would expect the rest sentences of this paragraph to be all about the solubility increase. The above stipulation was realized by a heuristic algorithm: given a pool of molecules with their property (in tabular form), we first clustered the molecules in terms of the molecular scaffolds with randomly selected clustering centers. If the similarity between a molecule and a center is greater than 0.6, the molecule is clustered at that center. The number of clustering centers would dynamically increase until all of the molecules in the pool are classified. 

Apart from the modification of the molecule to a single property, we also consider the compositions of multiple properties, which are mainly involved in the simple molecular properties that can be calculated by Python scripts. For example, we include LogP, Topological polar surface area (TPSA), and their composition in the training set for model training. In total, we collect over 25 million modification paragraphs and 2 billion molecules to build the training dataset. The dataset involves over 10, 000 different molecular properties, activities, and compositions. In addition to the SMILES molecule, we also added descriptions of the property optimizations in each paragraph to build the relationship between the molecule structures and the semantic meaning of the properties. 

\subsection{Group-based molecular representation (GMR)}
The core of the GMR framework involves decomposing molecules into structural groups and noting the inter-group connections, facilitating the group-based reconstruction of SMILES strings. GMR begins by assigning a unique string identifier to each structural group, thereby constructing a comprehensive dictionary. When a molecule is processed within the GMR framework, its SMILES string is converted into an encoded string that encapsulates both the identity of the structural groups and their positional relationships within the molecule. For computational analyses or property evaluations that require the original SMILES string, the encoded string can be decoded by applying a decoding algorithm. In the specific implementation of the GMR framework, there are three key steps:

 (1) \textbf{Dictionary construction}: Initially, we leverage the extensive molecular data resources available in the ChEMBL database. Using SMILES expressions, we extract information about the ring structures in the molecule, merge intersecting rings, and identify structural groups on the rings. For the nonring parts of the molecule, we employed a strategy of breaking all C-C bonds and treating the remaining molecular fragments as independent structural groups. This approach allows us to decompose all the groups of the molecule, assign a unique string identifier to each group, and construct a comprehensive dictionary.

\begin{algorithm}
\caption{Molecular encoding algorithm}
\label{alg:molecular_encoding}
\begin{algorithmic}[1]

\Procedure{EncodeMolecule}{molecule}
\State $unprocessedMolecularGroups \gets \text{splitIntoGroups} (molecule) $
\State $groupConnections \gets []$ 

\While{$\text{len} (unprocessedMolecularGroups) > 1$}
\State $remainingGroups \gets \emptyset$

\For{$currentGroup \in unprocessedMolecularGroups$}
\State $molecule, isProcessed \gets \text{ProcessGroup} (molecule, currentGroup, groupConnections)$
\If{not isProcessed} 
\State $remainingGroups \gets remainingGroups \cup \{currentGroup\}$
\EndIf
\EndFor

\State $unprocessedMolecularGroups \gets remainingGroups$
\EndWhile

\State $initialGroup \gets \text{first} (unprocessedMolecularGroups) $
\State $moleculeEncoding \gets \text{getDictionaryString} (initialGroup) $

\ForAll{$connectionItem \in groupConnections$}
\State $moleculeAtom \gets connectionItem[molConnectedAtom]$
\State $groupAtom \gets connectionItem[groupConnectedAtom]$
\State $group \gets \text{getDictionaryString} (connectionItem[group])$
\State $connectionEncoding \gets moleculeAtom + "/" + groupAtom + group$ 
\State $moleculeEncoding \gets moleculeEncoding + connectionEncoding$
\EndFor

\State \Return $moleculeEncoding$
\EndProcedure
\end{algorithmic}
\end{algorithm}

 (2) \textbf{Molecular encoding}: The encoding process, as depicted in Algorithm~\ref{alg:molecular_encoding}, is initiated by splitting the SMILES string of an individual molecule into several structural units. We systematically decompose the molecule, removing each structural group and verifying the connectivity of the molecule after each removal. If the molecule remains connected after removal of a structural group, we record the two atoms at the connection point. Since marking the connection point may cause changes in the atom index and affect the subsequent normalization process of the molecule, we use a breadth-first search algorithm to characterize the marked atoms and their adjacent area in detail. This forms a feature description of the atoms. After removing the mark and performing SMILES normalization on the structural group and molecular fragment after splitting, we use the previously obtained atom feature description to re-identify and record the atom index of these two atoms as connection location information. This process is repeated until the molecule is completely decomposed into a single structural group, at which point the molecular fragment serves as the starting point for encoding. Starting from the encoding starting point, we build the basis of the encoding string according to the corresponding string in the dictionary. Subsequently, we traverse the structural groups recorded during the removal process and their corresponding atom indices, using the “/” character to separate different location information. Gradually, we integrate the strings corresponding to the structural groups in the dictionary and the location information into the encoding string, eventually generating a complete and accurate molecular encoding result.

 (3) \textbf{Molecular decoding}: The decoding process is essentially the reverse of the encoding process. We use the encoded molecular fragment as the starting point for splicing and gradually recombine each structural group in the correct position according to the connection information recorded in the encoding. This process is repeated until all structural groups are correctly spliced back, thereby restoring the original molecular SMILES. This ensures the integrity and reversibility of molecular information.

\subsection{The implementations of DrugLLM}
Similar to ChatGPT, the training objective of DrugLLM is to iteratively predict the next token of the paragraphs. Formally, a generated paragraph $\bm x$ is composed of the optimization description $\bm o$ and the molecule modifications $\bm m$, given by
\begin{align}
\bm x= [\bm o, \bm m_1, \bm m_2, \cdots, \bm m_N], 
\end{align}
where $\bm m_n$ stands for the $n$-th case of the molecule modification. Essentially, DrugLLM is to approximate the probability 
\begin{align}
P (x_t|x_1, x_2, \cdots, x_{t-1}) = \text{DrugLLM} (x_1, x_2, \cdots, x_{t-1}), 
\end{align}
where $x_t$ is a token in the paragraph $\bm x$.

However, the testing objective of the few-shot molecule optimization is to predict the optimized molecules given the few shots of molecule modifications. That is
\begin{align}
m_g = \text{DrugLLM} (\bm m_1, \bm m_2, \cdots, \bm m_K, m_{o}), 
\end{align}
where $m_g$ stands for the generated molecules of DrugLLM in the $K$ shots input. $m_{o}$ represents the query molecule that needs to be optimized. Similarly, the testing objective of the zero-shot molecule optimization is to predict the optimized molecules given the descriptions of the optimization tasks, given by
\begin{align}
m_g= \text{DrugLLM} (\bm o, m_{o}).
\end{align}
 
We adopt the LLaMA model architecture as the basic backbone of our DrugLLM. Specifically, DrugLLM is a Transformer decoder with 32 layers and 32 attention heads. The hidden dimension is set to 4096 and the batch size is 64. As a result, DrugLLM has 7 billion parameters, all of which are updated during the pre-training. 
The training process employs the AdamW optimizer with a learning rate of $3\times 10^{-5}$, and we adopt a cosine annealing schedule to adjust the learning rate.

\subsection{The implementations of the competitive baselines}
There are very few methods that support few-shot molecule optimization or generation. In this work, we take the widely used and powerful molecule generators as baselines, including JTVAE, VJTNN, MoLeR, and a control model (random generation). For these methods, we used the released official codes and adapted them to perform few-shot optimization tasks. For JTVAE, which is designed to generate molecules from a pre-trained latent space, we follow the convention to optimize the molecules based on the released pre-trained JTVAE model. To accomplish the $K$-shot optimization, we leveraged the $K$ modification cases as the training samples and used the trained JTVAE to generate the new molecules for evaluation. The evaluation procedures of VJTNN and MoLeR are similar to JTVAE except for the exclusion of the pre-trained model. The random generation model is implemented by random sampling based on the latent space of the pre-trained JTVAE.

We further incorporated advanced large language models as benchmarks, including ChatGLM, ChatGPT, and GPT-4. These models require carefully constructed prompts to generate relevant output data. In crafting these prompts, we integrated the characteristic descriptions of the molecules to be optimized and instances of a few-shot optimization to ensure that the model can accurately understand the task requirements. For ChatGLM, we used the officially released code and pre-trained models, deployed and executed in a local server environment, to obtain the model prediction results. In contrast, for ChatGPT and GPT-4, we utilized their online service capabilities by transmitting the processed input data through their official API interfaces, thereby eliciting the corresponding model outputs from remote servers.

\section*{ACKNOWLEDGMENTS}
This work was supported by the National Natural Science Foundation of China (No. 62206192); the Natural Science Foundation of Sichuan Province (No. 2023NS-36 FSC1408); the Science and Technology Major Project of Sichuan Province; and the Fundamental Research Funds for the Central Universities (No. 1082204112364).

\bibliographystyle{dcu}
\bibliography{ref}

\end{document}